\documentclass[twocolumn,pra,letterpaper,superscriptaddress,floatfix]{revtex4}
\usepackage{graphicx,psfrag,bbm,latexsym,color,dcolumn,bm,dsfont,bbm,color,mathrsfs,bbold,latexsym,amsmath,amsfonts,amssymb,epsfig}

\newcommand{\beq}{\begin{equation}}
\newcommand{\eeq}{\end{equation}}

\newcommand{\beqa}{\begin{eqnarray}}
\newcommand{\eeqa}{\end{eqnarray}}

\newcommand{\A}{$\,\, \frac{3}{2}-\frac{3}{2}-1-1\,\,$}
\newcommand{\B}{$\,\, \frac{3}{2}-\frac{3}{2}-\frac{1}{2}-\frac{1}{2}\,\,$}
\newcommand{\C}{$\,\, \frac{1}{2}-\frac{1}{2}-1-1\,\, $}

\begin{document}

\title{Leading Corrections to Finite-Size Scaling for Mixed-spin Chains}
\author{R. Bischof}
\affiliation{
Institut f\"ur Theoretische Physik, Universit\"at Leipzig, D-04109 Leipzig, Germany.}
\author{P.R. Crompton}\affiliation{
Institut f\"ur Theoretische Physik I., Universit\"at Hamburg, D-20355 Hamburg, Germany.}\affiliation{
Institut f\"ur Angewandte Physik, Universit\"at Hamburg, D-20355 Hamburg, Germany.}
\affiliation{
Center for Theoretical Physics, Massachusetts Institute of Technology, Cambridge, MA  02139, USA.}
\vspace{0.2in}
\date{\today}

\begin{abstract}
{We identify the leading corrections to Finite-Size Scaling relations for the correlation length and twist order parameter of three mixed-spin quantum spin chains for the critical feature that develops at, $\theta=\pi$, corresponding to a change in the topological realisation of the groundstates.}
\end{abstract}

\maketitle

The effective model that describes the groundstate of quantum spin chains is the 2d O(3) model with a nonzero 
$\theta$-term \cite{Haldane}. Whilst it is clear that the groundstates of quantum spin chains can be related to the renormalization group fixed points of the SU(2) WZW model \cite{2}, it is less clear how the phase-space of just one nonperturbative system can be defined from the perspective of renormalization.  What we would like to understand, via the 2d O(3) model, is how the values of the couplings in the action (spin stiffness, spin wave velocity, etc.) influence the groundstate. This would tell us how the renormalization group flow of the SU(2) WZW can be related to quantum spin chains on a physical scale. However, the value of these couplings is dependent on the nonlocal fluctuations of the system, for which we either have to make some form of ansatz, or determine phenomenologically. Conversely, a Monte Carlo calculation is a very good way of generating a system with nonlocal fluctuations. Locally, the interaction couplings of numerical lattice system, such as a Quantum Monte Carlo simulation, are well-defined. We could therefore imagine relating the coupling constants of the 2d O(3) model to the lattice interaction couplings to quantify the nonlocal fluctuations.  However, although it is meaningful to vary the value of the couplings arbitrarily in numerical methods, not all of the corresponding physics is renormalizable as a continuum theory. The difficulty is that, $\theta$, cannot be varied continuously between, $\theta=0$, and, $\theta=\pi$, to describe the groundstate of the O(3) model. There are nonintegrable singularities in the model that prevent this \cite{poly}. 

Several numerical studies have been performed previously that identify the critical regions associated with
the topological rearrangement of the groundstate in quantum spin chains. This critical behaviour occurs (in the absence of external fields) when, $\theta$, changes value between, $\theta=0$, and, $\theta=\pi$, via changes in the nonperturbative vacuum \cite{zL}-\cite{n=4}. What has not been identified in these previous numerical studies is whether or not the nonperturbative behaviour constitutes a continuous topological phase transition in, $\theta$. In principle there can be a second order transition at zero temperature in quantum spin chains without contradiction with the important Mermin-Wagner theorem \cite{MW}. This theorem excludes the development of long-range order in finite temperature quantum spin chains, and therefore the development of a genuinely massless phase corresponding to a broken symmetry, but not at zero temperature.  What we are able to do via Quantum Monte Carlo analysis is to keep the volume of the lattice system fixed in physical units, but to vary the size of the nonlocal fluctuations by changing the number of lattice sites in the physical volume. The purpose of this new study is therefore to try and quantify the renormalization group flow around, $\theta=\pi$, in terms of nonlocal fluctuations, to define the physical scale of the renormalization group flow.
\section{mixed-spin model}
In this article we investigate the nature of the groundstate critical feature corresponding to, $\theta= \pi$, by studying chain systems that are amenable to numerical evaluation: ones for which we can circumvent the numerical complex-action problem \cite{meron}. Specifically, we consider AFM mixed chains for the Hamiltonians with spins ${\bf S}^{\bf{a}}$ and ${\bf S}^{\bf{b}}$ arranged in cells of period-4. We choose without loss of generality $J_{\bf{aa}} \equiv J_{\bf{bb}}\equiv 1$, and the coupling anisotropy $\alpha \equiv J_{\bf{ab}} / J_{\bf{aa}}$, $L$ being the chain length. We vary the coupling interaction strength between spins to effectively vary, $\theta$, through, $\alpha$. 

\beqa
\label{model}
  H &=& \sum_{j=0}^{L/4-1} \,
  ( J_{\bf{aa}} \, {\bf S}^{\bf{a}}_{4j} \cdot {\bf S}^{\bf{a}}_{4j+1}
  + J_{\bf{ab}} \, {\bf S}^{\bf{a}}_{4j+1} \cdot {\bf S}^{\bf{b}}_{4j+2}
\nonumber \\
  &+& J_{\bf{bb}} \, {\bf S}^{\bf{b}}_{4j+2} \cdot {\bf S}^{\bf{b}}_{4j+3}
  + \,\, J_{\bf{ab}} \, {\bf S}^{\bf{b}}_{4j+3} \cdot {\bf
S}^{\bf{a}}_{4j+4} ),
\eeqa

The periodicity in (\ref{model}) is a necessary condition for the constraint equation used to define the groundstate given in terms of a nonlocal singlet operator \cite{LSM}\cite{Haldane}. A detailed picture of all possible realisations of these singlets for mixed-spin chains is given in \cite{DMRG}. Specifically, if three different couplings are used for the periodic cell then it is possible for these singlets to be realised between next-to-nearest-neighbour sites, but here we restrict ourselves to the case of two AFM couplings and consequently have nearest-neighbour singlets. Our reason for looking at mixed-spin chains, rather than the more generalised bond-alternating chain \cite{BA}, is that there are differences in these groundstate parametrisations, as a function of spin magnitude, that allows us to identify important nonlocal trends in the numerics. In these groundstate parametrisations the nonperturbative condition, $\theta = \pi$, arises when the singlet source term is vanishing \cite{Takano}. At this point for a bond-alternating chain,
\beq
\frac{\theta}{2 \pi} = 2s_a J_{\bf{ab}} / (J_{\bf{aa}}+J_{\bf{ab}}),
\eeq
where, $s_{a}$, is the spin magnitude, and for a mixed-spin chain,
\beq
\label{eff2}
\frac{\theta}{2 \pi} = 2s_as_b (s_a+s_b) /
(s_a^2+s_b^2+2s_as_bJ_{\bf{aa}}/J_{\bf{ab}}).
\eeq
In the first case, for bond-alternating chains, $J_{\bf{ab}}/J_{\bf{aa}} = 1$ is always a solution for half-integer systems, and for integer systems the critical ratio of, $J_{\bf{ab}}/J_{\bf{aa}}$, tends to zero with increasing, $s_a$. In the second mixed-spin case the critical ratio of couplings always differs significantly at successive critical regions within a given model, at $\theta=\pi, 3\pi, 5\pi ...\,,$ etc.. The value of, $\theta$, associated with critical behaviour therefore has an implicit dependence on the spin magnitude for mixed-spin chains, 
which is not as evident for bond-alternating chains.

Although in these treatments the groundstate parametrisations can be used to identify the number of critical points in, $\theta$, these parametrisations cannot be used to properly define local expansions in, $\theta$, about the critical point at, $\theta=\pi$, in physical units. The reason for this is that a simple ansatz has to be defined in order to be able to quantify the nonlocal fluctuations of the vacua. However, by definition this is a nonperturbative quantity, because the point,  $\theta = \pi$, is asymptotically free \cite{poly}. Consequently there are major differences between the value of the critical point in physical units in the numerics \cite{zL}-\cite{n=4} and these analytic treatments \cite{Takano}, because the scale of the nonlocal fluctuations is essentially different in each case. To resolve this discrepancy we really need to parametrise the scale dependence of the nonlocal fluctuations in terms of physical units.

\section{nonperturbative renormalization Group Flow}
Recently, what has been proposed is that the nonlocal fluctuations lead to an additional term in the action of the corresponding conformal model, in the vicinity of, $\theta=\pi$, \cite{cont1}\cite{cont2}. This gives rise to a double Sine-Gordon model description of the nonlocal fluctuations. The original Sine-Gordon model has been presented in a lattice formalism in \cite{nakamura1}\cite{nakamura2}. As usual, when the topological Berry-phase term vanishes at, $\theta=\pi$, the action is described by the U(1) Gaussian theory given by the pure Sine-Gordon model.  In terms of the lattice formalism the renormalization group equations of this original Sine-Gordon model is given by,

\beq
\label{RG}
\frac{da_{s}}{dl} = -\frac{1}{2} a_{s}^{2}\left(\frac{a_{s'}}{\pi}\right)^{2}, \quad
\frac{da_{s'}}{dl}=a_{s'}(2-2a_{s})
\eeq

where, $l={\rm{log}} L$, and, $a_{s}$, and, $a_{s'}$, are nonperturbative or phenomenologically determined parameter, which relate respectively to the couplings of the Berry-phase term and the kinetic terms. In the original Sine-Gordon model, these relations in (\ref{RG}) imply that there is a renormalization of the gap in the vicinity of the critical point. This has be treated explicitly by via perturbative expansion in \cite{saus}. The difference for the double Sine-Gordon model is that the treatment has two terms corresponding the topological phase, which have different prefactors. Thus, in addition to the renormalization of the gap, the inclusion of nonlocal fluctuations will lead to the topological Berry-phase being effectively made anisotropic via a Lorentz boost. The leading correction to (\ref{RG}) should then be a relatively simple factor, which accounts for this effective anisotropy. For the second order scaling in (\ref{RG}) this anisotropy will mean a logarithmic correction to the free energy, defined in lattice units. 

Of course, both linear and logarithmic corrections to FSS can arise in the pure Sine-Gordon model from conformal symmetries. Quantum systems inherently have a conformal charge, $c$, that can be derived from their associated Virasoro algebra \cite{con}\cite{8}. For the Gaussian system at the renormalization group fixed point (with $c=1$) this will lead to an additional term in the free energy with an explicit temperature dependence. Similar universal correction contributions can also arise from the topology of the Lattice system: in \cite{Fisher} it was argued that boundary conditions can lead to linear corrections to FSS from the conformal block. These are all potential independent sources for corrections to scaling, from the source of scaling we wish to focus on. From Haldane's conjecture all quantum spin chains mapping to the point, $\theta=\pi$, should share these universal conformal corrections and be of the same universality class. In this article, however, we are specifically interested to identify behaviour relating to the extra term in \cite{cont1}\cite{cont2}. The contributions coming from this term are nonuniversal, being dependent on the nonlocal fluctuations of, $\theta$. For the mixed-spin models we are considering, however, the critical anisotropy, $\alpha$, is strongly dependent on spin magnitude for the different transition between topological sectors (\ref{eff2}), even though, $c=1$, for all of these critical points. A strong spin magnitude dependence, and logarithmic corrections, should therefore be indicative of the nonuniversal FSS correction due to the nonlocal fluctuations in our study.

\section{Numerical Measurements}

We consider the FSS behaviour of two thermodynamic indicators: the correlation length, evaluated via a second-moment estimator method \cite{xi}, and the twist order parameter \cite{nakamura1}. The second-moment estimator for the correlation length, $\xi$, is defined through the Fourier transform of a series, and potentially, this series 
is divergent away from zero temperature. Also, since we expect the critical behaviour to be of the SU(2) WZW  
universality class, potential systematic errors are indicated from analytic results for half-integer chains at finite
temperatures \cite{temperature}. These results identify the presence of logarithmic corrections to the finite-temperature scaling relations, although, this should not effect our identification of nonuniversal corrections to FSS. The twist order parameter, $z_{L}$, can be considered as a generalisation of the string order parameter to systems of broken translational invariance, and it is an exact order parameter for the groundstate singlet \cite{nakamura1}. The asymptotic form of, $z_L$, is given by,
\beq
\label{twist}
z_L =
  \langle\Psi{(n)}|U|\Psi^{(n)}\rangle
  =(-1)^n[1-{\cal O}(1/L)],
\eeq
where, $n$, counts the number of groundstate singlets, $\Psi$, of a given topology per periodic cell, and the operator, $U$, is defined by,
\beq
\label{operator}
U\equiv\exp\left[{\rm i} \frac{2\pi}{L}\sum_{j=1}^L j S_j^z\right],
\eeq
where, $S_j^z$, is the $z$-component of spin at lattice site, $j$, \cite{zL}. This order parameter changes sign as it passes through the critical topological region. From the asymptotic definition in (\ref{twist}) we may expect generic linear corrections to, $z_{L}$, that potentially affects the FSS. However, this again should not prevent us from identifying the presence of non-universal corrections to FSS.

\section{Results and Analysis}
We have investigated three mixed-spin chain systems with the periodic cells \A, \B and \C which we label $A$, $B$ and $C$. Fits to the FSS trial functions were extracted over the numerically accessible temperature regimes for each spin system. Below the smallest temperatures presented in the tables we found that the continuous-time Quantum Monte Carlo method we used became inefficient to update, and above the largest temperatures the data became too smooth to fit. This is consistent with the Sine-Gordon model renormalization group flow in (\ref{RG}), which indicates that above a certain temperature scale the flow of the gap should go to a finite value above the gapless point. From the possible topological realisations of the groundstate identified in \cite{DMRG}\cite{Takano}, we confirmed that model $A$ has two critical regions, whilst $B$ and $C$ have each one, although there is strong quantitative disagreement about the location of these transitions. We believe this is due to the renormalization scale dependence of the nonlocal fluctuations, which we now explicitly treat in this analysis. For our numerical measurements we generated 100,000 independent configurations per Lattice ensemble for system sizes $L$=16, 32, 64, 128, 256, and 512. The critical value of the coupling anisotropy, $\alpha_{c,L}$, was determined on finite chain lengths from interpolation using the Levenberg-Marquadt method to find the maxima of, $\xi$, for the correlation length, and the point at which, $z_{L}=0$, for the twist order parameter. 

Our basic assumption is that the scale of the nonlocal fluctuations scales linearly with spatial length in the method we have used. We verify this in Fig.1, where we plot the length dependence of the maxima of, $\xi$. When either the inverse temperature, $\beta$, is increased at fixed chain length, $L$, or vice versa there is a departure from linear (second order) scaling when either scale approaches the magnitude of the correlation length. However, with fixed ratio between the two scales the correlation length can always be made to scale linearly with, $L$. Therefore there is a direct correlation between the chain length and the size of the system in physical units of, $\beta$.
\begin{figure}
\epsfxsize=3.3 in\centerline{\epsffile{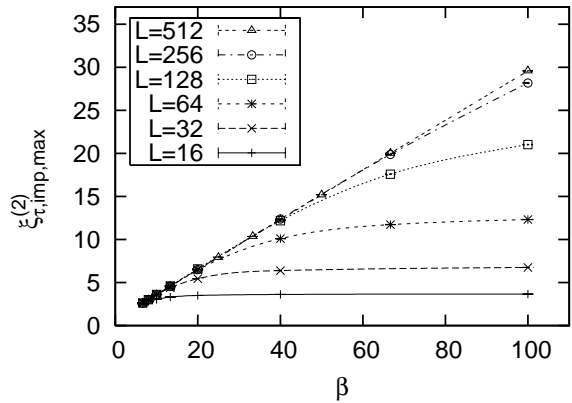}}
\caption{Correlation length maxima, $\xi_{max}$, versus inverse temperature,
$\beta$, as a function of chain length, $L$.}
\end{figure}

We extracted the shift exponent, $\lambda$, via fits to,
\beq
\label{fit1}
\alpha_{c,L} - \alpha_{c} =  a/L+b/L^{\lambda},
\eeq
where, $\alpha_{c}$, $a$, $b$, and, $\lambda$, are fitted constants. The results are presented in Tables 1 and 2. To summarise: no stable fit was obtained for the correlation length measurement of the second transition point in $A$ which corresponds to, $\theta=3\pi$, and the exponents all appear highly nonuniversal. We anticipate both universal linear and nonuniversal logarithmic corrections to scaling due to, respectively, the conformal symmetries and the nonlocal fluctuations in our analysis. Although we can generate lattice ensembles to correspond to a range of different nonlocal fluctuation scales, it is difficult for us then to make a direct comparison between nonuniversal and universal corrections. The problem is that in the continuous-time method varying the scale of the nonlocal fluctuations also changes the renormalization scale of the gap, from (\ref{RG}), because the size of the lattice is defined in physical units, rather than number of lattice sites. Although the size of the conformal correction should be fixed, the lattice volume is fluctuating. The way we treated this was following \cite{sphere}, to look at the relative strengths of universal and nonuniversal corrections. Since the lattice units are nonuniversal, we instead fix the nonuniversal exponent. The nonuniversal exponents, from Tables 1 and 2, range between 1.4-3.1, and so we fix the value of the exponent to be, $\lambda=2$, to make the comparison, via the constrained fit,
\beq
\label{fit2}
\alpha_{c,L} - \alpha_{c} = a/L+b\log(L)/L^{2}.
\eeq

\begin{table}
\begin{center}
\begin{tabular}{|l|l|r|r|}
\hline
$T$        &         & $\alpha_{c}$ & $\lambda$  \\
\hline
\hline
0.01       & $\xi$   & 0.48224(15)  & 1.23(24)   \\
           & $z_{L}$ & 0.48360(48)  & 1.32(5)    \\
           & $z_{L}$ & 1.3126(14)   & 1.41(9)    \\
\hline
0.025      & $\xi$   & 0.48279(21)  & 1.35(32)   \\
           & $z_{L}$ & 0.48240(43)  & 1.31(4)    \\
           & $z_{L}$ & 1.3146(7)    & 1.47(6)    \\
\hline
0.04       & $\xi$   & 0.48314(18)  & 1.51(17)   \\
           & $z_{L}$ & 0.48189(47)  & 1.25(4)    \\
           & $z_{L}$ & 1.3146(5)    & 1.38(6)    \\
\hline
\end{tabular}
\end{center}
\caption{Model $A$, \A. }
\end{table}

\begin{table}
\begin{center}
\begin{tabular}{|l|l|r|r|l|r|r|}
\hline
$T$        &         & $\alpha_{c}$ & $\lambda$ & T & $\alpha_{c}$ & $\lambda$ \\
\hline
\hline
0.01       & $\xi$   & 0.62132(5)   & 2.32(47)  & 0.01 & 0.76244(5)   & 2.72(14)  \\
           & $z_{L}$ & 0.62107(49)  & 1.59(5)   &      & 0.76246(17)  & 1.7(1)   \\
\hline
0.025      & $\xi$   & 0.62125(17)  & 1.62(47)  & 0.025 & 0.76289(8)   & 3.1(3)   \\
           & $z_{L}$ & 0.62141(17)  & 1.57(4)   & 	& 0.76389(17)  & 1.55(6)   \\
\hline
0.05       & $\xi$   & 0.62184(33)  & 1.32(29)  & 0.04	& 0.76440(13)  & 2.78(38)  \\
           & $z_{L}$ & 0.62256(52)  & 1.44(4)   & 	& 0.76508(29)  & 1.58(7)   \\
\hline
0.075      & $z_{L}$ & 0.62211(65)  & 1.48(5)   & 0.05	& 0.76526(21)  & 2.41(33)   \\
\hline
\end{tabular}
\end{center}
\caption{Models $B$, \B, and $C$, \C.}
\end{table}

The qualitative comparison of the linear, $a$, and logarithmic corrections, $b$, is presented for the correlation length of models $A$, $B$ and $C$ in Fig.2. From our fits, $a$, is significantly different from zero for $B$ and $C$, and in addition there is an unexpected temperature dependence for $B$. Similar behaviour is seen for, $b$, although 
the size of correction is considerably larger, supporting our premise of a leading logarithmic correction. Since the temperature is fixed in physical units for each measurement, this unforeseen temperature dependence is most likely due to the systematic errors in the improved estimator method for the correlation length, since it is largely absent for the corresponding Twist order parameter plots in Fig.3. The FSS corrections are strongly dependent on the spin magnitudes for the specific models, $A$, $B$ and $C$, which also suggests that the leading corrections to scaling at the, $c=1$, point are nonuniversal. The Twist order parameter measurements of Fig.3 are generally more stable than the corresponding correlation length measurements, and we are able to more distinctly see the large separation in values between the size of the linear and logarithmic corrections. The twist order parameter is perhaps a more consistent with our expectations, because it is a more direct measure of the singlet state than the correlation length second-moment improved estimator.

\begin{figure}
\epsfxsize=3.3 in\centerline{\epsffile{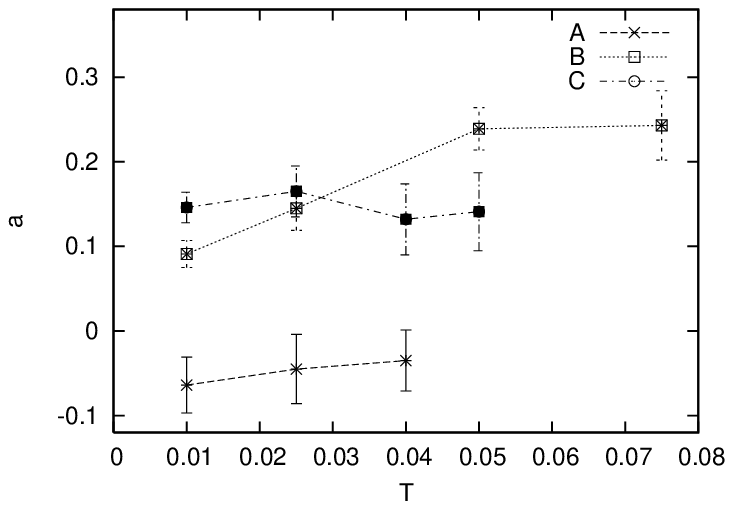}}
\epsfxsize=3.3 in\centerline{\epsffile{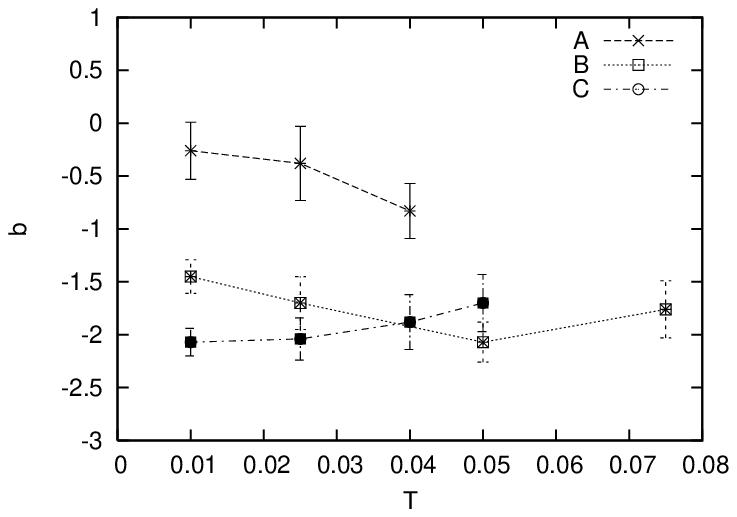}}
\caption{Linear correction term, $a$, and logarithmic correction term, $b$, for the correlation length, $\xi$, versus temperature, $T$. }
\end{figure}

\begin{figure}
\epsfxsize=3.3 in\centerline{\epsffile{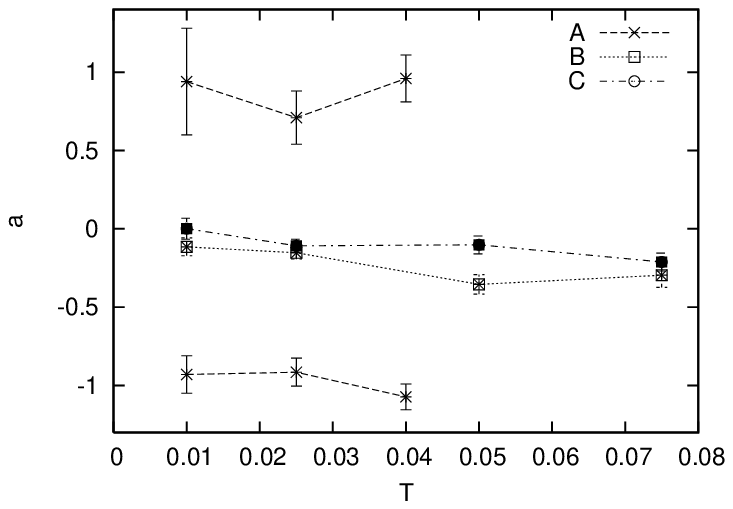}}
\epsfxsize=3.3 in\centerline{\epsffile{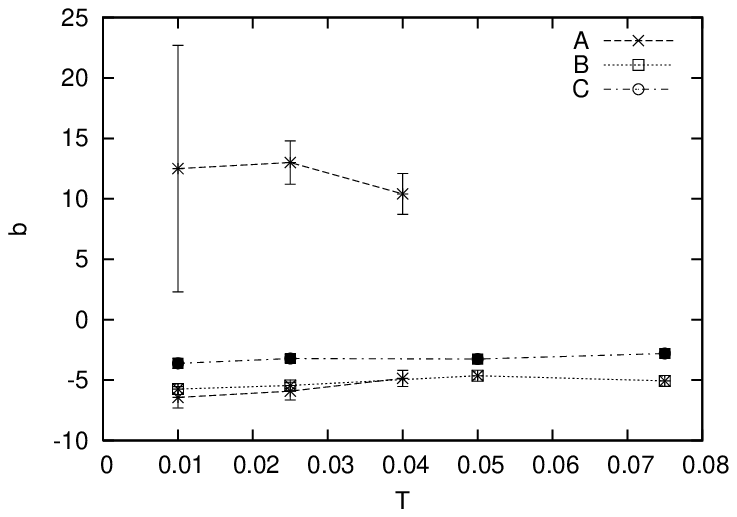}}
\caption{Linear correction term, $a$, and logarithmic correction term for, $b$, for the Twist order parameter, $z_{L}$, versus temperature, $T$.}
\end{figure}

\section{Summary}

In this article we have identified the leading corrections to FSS of three mixed-spin chains, via Quantum Monte Carlo 
analysis, in the vicinity of the critical point at, $\theta=\pi$. The scaling picture presented is one defined in terms of, $\beta$, and the effective value of, $\theta$, determined from the fluctuations of the lattice units, and the translationally varying interaction coupling, $J_{ab}$. We have identified that the leading corrections are logarithmic and nonuniversal, dependent on both the spin magnitude of the system and the nonlocal fluctuations of the lattice ensemble. Measurements have been compared between a second-moment estimator measurement for the correlation length, and the twist order parameter, and it was found that the latter were the more numerically stable. Our numerics confirm the effective renormalization region picture developed in \cite{cont1}\cite{cont2} of the addition of nonlocal fluctuations to the scaling picture of the $c=1$ conformal treatment. However, it is not clear that we can determine the dependencies on, $\theta$, and, $\beta$, from our measurements sufficiently to be able to define the corresponding lattice $\beta$-function. The reason for this is that the continuous-time method we have used is defined in fixed physical units, but with a fluctuating space-Euclidean time volume. Consequently, the lattice units of our FSS are phenomenological, and it is difficult to obtain an unrenormalized value for the central charge. It is therefore difficult for us to make a comparison of the unrenormalized nonuniversal and universal corrections of the system, since the lattice units are themselves nonuniversal. However, we have been able to make a qualitative comparison of the size of the universal and nonuniversal correction terms by picking an arbitrary common renormalization scale for the scaling functions. Our numerical results for the nonuniversality indicate that the contribution from different topological sectors, defined in terms of the topologically distinct groundstate realisations, is crucial to understanding the renormalization group flow of numerical Quantum Monte Carlo data. Although our specific analysis has been performed for quantum spin chains in the vicinity of the zero temperature fixed point of the double Sine-Gordon model, we anticipate our findings may be of general applicability to low-dimensional quantum spin systems, given the ubiquitous character of the O(3) model in describing topological quantum effects.

We would like to acknowledge the financial support of the DFG through the SFB 668 program, and also that of the EU Marie Curie 'Host Development' program.

\end{document}